\documentclass[pra,twocolumn,showpacs]{revtex4-1}
\usepackage{}
\usepackage{savesym}
\usepackage{amsmath}
\savesymbol{iint}
\usepackage{txfonts}
\restoresymbol{TXF}{iint}
\usepackage{amssymb}
\usepackage{mathrsfs}
\usepackage{bbm}
\usepackage{bm}
\usepackage{graphicx}

\newcommand{\ket}[1]{|#1\rangle}
\newcommand{\bra}[1]{\langle#1|}

\begin{document}

\title{Nonadiabatic geometric rotation of electron spin in a quantum dot by 2Pi hyperbolic secant pulses}
\author{Pei Pei}
\email{ppei@mail.dlut.edu.cn}
\author{Feng-Yang Zhang}
\author{Chong Li}
\author{He-Shan Song}
\email{hssong@dlut.edu.cn}
\affiliation{School of Physics and Optoelectronic Engineering,\\ Dalian University of Technology, Dalian 116024, China}
\date{\today}
\begin{abstract}
In this paper, the geometric and dynamic phase components of overall phase induced by $2\pi$ hyperbolic secant pulses in a quantum dot is analyzed. The dependence of two phase components on the ratio of the Rabi frequency to the detuning is investigated. Numerical results indicate that only for one resonant pulse the induced overall phase is purely the geometric phase. With other values of the ratio the overall phase consists of a nonzero dynamic part. The effect of spin precession to decrease the dynamic phase is characterized and discussed by analytical and numerical techniques. Utilizing the symmetry relations of the phases, a scheme to eliminate the dynamic phase by multipulse control is proposed. By choosing the proper parameter for each pulse, the dynamic phases induced by different pulses cancel out. The total pure geometric phase varies from $-\pi$ to $\pi$, which realizes the arbitrary geometric rotation of spin. Average fidelity is calculated and the effects of magnetic field and decay of the trion state are compared and discussed. The results show the crucial role of weak magnetic field for high fidelity (above $99.3\%$).
\end{abstract}
\keywords{ }
\pacs{78.67.Hc, 03.67.Lx} \maketitle

\section{Introduction}
Optical control of a single electron spin in a quantum dot (QD) is a key ingredient for implementing quantum information processing (QIP) in a scalable solid-state system \cite{Imamoglu99}. In recent years, great efforts have been made to the physical implementation of such optical approach. It has been experimentally demonstrated using oscillating magnetic field generated by radio-frequency (RF) pulses on timescales of nanoseconds \cite{Koppens06,Koppens07}. However, ultrafast optical technique is more attractive due to its ability of enabling spin rotations to be completed on a picosecond timescale which is much shorter than spin coherent times \cite{Greilich06}, leading to abundant theoretical proposals \cite{Kis02,Troiani03,Calarco03,Chen04,Emary07,Economou06,Economou07,Dutt06,Clark07} and several experimental achievements \cite{Berezovsky08,Fu08,Press08,Greilich09,Kim10}.

The proposal $\mathcal{T}$ proposed by Economou \textit{et al}. \cite{Economou06,Economou07} is based on the attractive feature of the fast laser pulse with a hyperbolic secant temporal envelope \cite{Rosen32}. After a $2\pi$ sec pulse, the two-level system of a QD completes a Rabi oscillation and the population returns to its initiate state while having acquired an overall phase. This initiate state with the overall phase and a third state which does not couple to the laser pulse span the qubit space. The fast spin rotations about the optical axis is thus achieved, and the rotation angle is determined by the ratio of the Rabi frequency to the detuning, $\Omega/\Delta$ \cite{Economou06}. In general, the $2\pi$ Rabi oscillation may be considered as a nonadiabatic cyclic evolution, therefore the overall phase consists of a dynamic component and a geometric component defined by Aharonov and Anandan \cite{Aharonov87}. Particularly, when the rotation angle $\phi=\pi$ ($\Omega/\Delta\rightarrow\infty$), i.e. the resonant case, the dynamic phase component is zero and the overall phase is exactly equals to the geometric phase component. The geometric phase depends only on some global geometric features e.g. the curve in the parameter space, not on the duration of interaction. Besides, geometric phases may be robust against dephasing. Although this property is not predicted in general systems \cite{Tong04}, it manifests in particular open systems where precession is in the equatorial plane of the Bloch sphere \cite{Carollo03}. Therefore the quantum computation based on pure geometric phase, so called geometric quantum computing (GQC) \cite{Zanardi99,*Pachos99}, may have inherent error-resilient advantage. Proposal $\mathcal{T}$ is thus one of the methods to physically implement GQC. However, the cycle-averaged expectation value of the Hamiltonian is not always zero with various values of the ratio $\Omega/\Delta$, leading to a nonzero dynamic component and diverse geometric component. In this paper, we will analyze the geometric and dynamic phase components induced by the $2\pi$ sec pulse in detail, through investigating the dependence of the two components on the spin rotation angle and on the ratio $\Omega/\Delta$. Two methods are utilized: one is to calculate the phases components by substituting the state vector with analytic solutions of Ref.~\cite{Economou06} in intermediate steps, the other is full numerical simulation. Besides, the discrepancy of two results is discussed and the effect of the other interaction mechanism within the system i.e. the spin precession is demonstrated.

To generalize the proposal $\mathcal{T}$ to universal nonadiabatic geometric quantum gates, e.g. the arbitrary angle rotation about an axis is purely the geometric phase, one needs to avoid or remove the dynamic component when $\phi\neq\pi$ ($\Omega/\Delta\nrightarrow\infty$). In previous studies \cite{Wang01,*Wang02E,Wang02,Li02,Solinas03,Zhu02PRL,*Zhu02PRLE,*Zhu02PRA,*Zhu03}, one method is to choose the dark states as the cyclic states, thus the dynamic phase component is always zero. This scheme has been proposed for nonadiabatic GQC with NMR \cite{Wang01,*Wang02E}, Josephson-junction \cite{Wang02}, trapped ions \cite{Li02}, and quantum dot systems \cite{Solinas03}. The other method is the multiloop scheme generalized from the adiabatic evolution case \cite{Falci00,Jones00}: let the system undergoes evolution along several closed loops, thus the dynamic phases accumulated in different loops may be canceled, while the geometric phase being added. This scheme has been demonstrated in Josephson junctions and NMR systems \cite{Zhu02PRL,*Zhu02PRLE,*Zhu02PRA,*Zhu03}. In this paper, we propose a scheme to eliminate the dynamic phases in quantum dot system. Though being similar with the multiloop method, we do not employ the RF pulses to generate the oscillating magnetic field in our scheme, but employ several picosecond sec pulses proposed in the proposal $\mathcal{T}$. The dynamic phases induced by different pulses cancel out, leading to arbitrary spin rotation with pure geometric phase.

The paper is organized as follows: in Sec. II, we give a brief review of the proposal for fast optical rotations of an electron spin trapped in a quantum dot, introduce the analytic solutions and calculate the expression of the phase components. In Sec. III, we present numerical results, analyze the phase components and demonstrate the effect of the spin precession to the dynamic phase. In Sec. IV, the scheme to eliminate dynamic phases by multipulse control is proposed. In Sec. V, we calculate the fidelity and discuss the effects of magnetic field and decay of the trion state. A summary and some prospects are provided in Sec. VI.

\section{Review of the proposal and calculation of the phase components}
\begin{figure}[htbp]
\includegraphics[width=8cm]{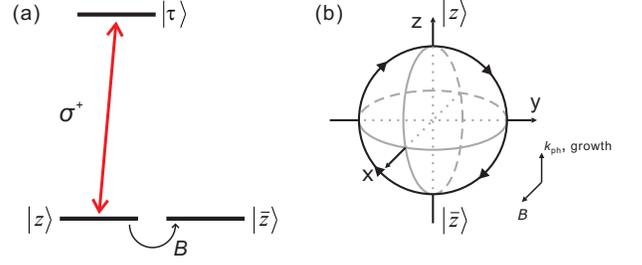}
\caption{\label{fig:1}(Color online) (a) Level diagram of the $\Lambda$ system in QD. In the Voigt geometry, two lower states are $\ket{z}$ and $\ket{\bar{z}}$ with electron spins in the $z$ direction, coupled by the magnetic field $B$. The intermediate state $\ket{\tau}$ is the trion state with the heavy spin pinned along $z$ axis. Circularly polarized light propagates parallel to the QD growth direction. Due to selective rules, $\sigma^{+}$ pulse only couples $\ket{z}$ and $\ket{\tau}$. (b) Bloch sphere sketches electron spin and precession of the spin around the magnetic field. The clockwise arrows mark the spin precession direction.}
\end{figure}

The nanostructure employed in the proposal $\mathcal{T}$ consists of arrays of self-assembled (In,Ga)As/GaAs QDs, each containing on average a single electron \cite{Yugova07}. In the Voigt geometry, the external magnetic field (along the $x$ axis) $B=0.29$T \cite{Greilich09}, applied perpendicularly to the QD growth direction ($z$ axis) which is parallel to the optical axis, as shown in Fig.~\ref{fig:1}(b). The two lower states $\ket{z}$ and $\ket{\bar{z}}$ with spins in the $z$ direction are coupled through the magnetic field, and they are superpositions of spin energy eigenstates in the $x$ direction. When the QD absorbs a photon, it's excitated to a trion state which consists of a singlet pair of electrons and a heavy hole. The hole spin is pinned along the growth direction due to strong confinement and spin-orbit interaction. Selection rules determine that specific circularly polarized light only couples one spin state to one trion state, e.g. $\sigma^{+}$ light couples $\ket{z}$ to $\ket{\tau}$, leading to the $\Lambda$-type system shown in Fig.~\ref{fig:1}(a). The Hamiltonian of system in the interaction picture reads
\begin{equation}
H=\hbar\left[
    \begin{array}{ccc}
      0 & \omega_{B} & 0 \\
      \omega_{B} & 0 & \Omega(t)e^{-i\Delta{t}} \\
      0 & \Omega(t)e^{i\Delta{t}} & 0 \\
    \end{array}
  \right]
,\label{H}
\end{equation}
with the basis $\{\ket{\bar{z}},\ket{z},\ket{\tau}\}$. $\omega_{B}$ is spin Larmor precession frequency and $\Delta$ is the detuning. $\Omega(t)=\Omega\textrm{sech}(\eta{t})$ is the time dependant Rabi frequency with a hyperbolic secant temporal envelope, where $\eta$ is the pulse bandwidth.

To obtain analytic solutions, the slow precession approximation \cite{Economou06} is required: when the Zeeman splitting is much smaller than the pulse bandwidth, i.e. $\omega_{B}\ll\eta$, the spin precession can be neglected within the pulse action. In this case the $\Lambda$-type system can be considered a direct sum of 1-dimension and 2-dimension systems, with the basis $\{\ket{\bar{z}}\}$ and $\{\ket{z},\ket{\tau}\}$ respectively. By the approach of Rosen and Zener \cite{Rosen32}, the second order equation of probability amplitudes is transformed to the hypergeometric differential equation through change of variable $z=(1/2)[\tanh(\eta{t})+1]$. Considering the spins are initialized in the $z$ direction, $C_{\bar{z}}(-\infty)=0,~C_{z}(-\infty)=1,~C_{\bar{z}}(-\infty)=0$, the analytic expression of state vector is obtained as
\begin{equation}
\ket{\psi(t)}=F(a,-a;c;z)\ket{z}-\frac{ia}{c}z^{c}F(a+c,-a+c;1+c;z)\ket{\tau}\label{psi},
\end{equation}
where $F$ denotes the Gauss hypergeometric series $_{2}F_{1}$ \cite{Abramowitz72} and the parameters $a=\Omega/\eta,~c=(1/2)(1+i\Delta/\eta)$.

If and only if $\Omega=\eta$, the pulse area $2\int\Omega{\textrm{sech}}(\eta{t})$ equals to $2\pi$. In this case during the pulse action the system undergoes a cyclic evolution between $\ket{z}$ and $\ket{\tau}$. After the pulse the population returns to $\ket{z}$ with an overall phase $\phi$:
\begin{eqnarray}
\ket{\psi(\infty)}=e^{i\phi}\ket{z},\\
\phi=2\arctan(\frac{\Omega}{\Delta}).\label{phi}
\end{eqnarray}
Considering the qubit space spanned by $\{\ket{\bar{z}},\ket{z}\}$, the unitary operator induced by the $2\pi$ sec pulse \cite{Economou06} is
\begin{equation}
U_{spin}\simeq{e^{i\phi/2}}\left[
    \begin{array}{cc}
      e^{-i\phi/2} & 0 \\
      0 & e^{i\phi/2} \\
    \end{array}
  \right],
\end{equation}
thus the spin rotation about $z$ axis is achieved.

Now we demonstrate the calculation of the phase components. The dynamic phase $\alpha$ is the cycle-averaged expectation value of the Hamiltonian and the geometric phase $\gamma$ is the rest part of removing the dynamic component from the overall phase \cite{Aharonov87}:
\begin{eqnarray}
\alpha&&=-\frac{1}{\hbar}\int_{-\infty}^{\infty}dt\bra{\psi(t)}H\ket{\psi(t)},\label{dynamic}\\
\gamma&&=\phi-\alpha.
\end{eqnarray}
One method (method I) to calculate the phase components is to utilize the analytic expression of $\ket{\psi(t)}$ under condition $\Omega=\eta$. We insert Eq.~(\ref{psi}) into Eq.~(\ref{dynamic}) and rewrite the dynamic phase as
\begin{widetext}
\begin{equation}
\alpha=\frac{\Omega^{2}}{\Delta^{2}+\Omega^{2}}\int_{-\infty}^{ \infty}\textrm{sech}^{2}(\Omega{t})[e^{-i\Delta{t}}(1-\tanh(\Omega{t}))^{\frac{-i\Delta}{2\Omega}}
(1+\tanh(\Omega{t}))^{\frac{i\Delta}{2\Omega}}(\Delta-i\Omega\tanh(\Omega{t}))+c.c.].\label{dynamic1}
\end{equation}
\end{widetext}
In this case the overall phase is calculated following Eq.~(\ref{phi}). We note that the expression of Eq.~(\ref{dynamic1}) is independent of $\omega_{B}$, resulting from the slow precession approximation. The other method (method II) is full numerical simulation, where the state vector is calculated through numerical solution of Schr\"{o}dinger equation. For this case the spin precession is taken into account. The specific considered parameters are demonstrated in Section V.

The integral interval is infinity in Eq.~(\ref{dynamic})(\ref{dynamic1}), which is in theory for including the whole sec pulse. In experiments the stimulating effect of pulses concentrates within the pulse duration, which is as short as 1.5ps \cite{Greilich09}, and the pulse tail on a long timescale can be safely neglected. On the other hand, especially for the full numerical simulation method where no slow precession approximation is made: if the time interval is too large, the population transferred to $\ket{\bar{z}}$ due to spin precession is prominent, thus the evolution of system can not be considered cyclic and the proposal $\mathcal{T}$ would not be ensured. Therefore in numerical simulation, we focus on the evolution of system during the pulse action. For both methods the integral interval is finite, with the same order of magnitude of the pulse duration.

\section{Analysis of the phase components and effect of the spin precession to the dynamic phase}
\begin{figure}[htbp]
\includegraphics[width=8cm]{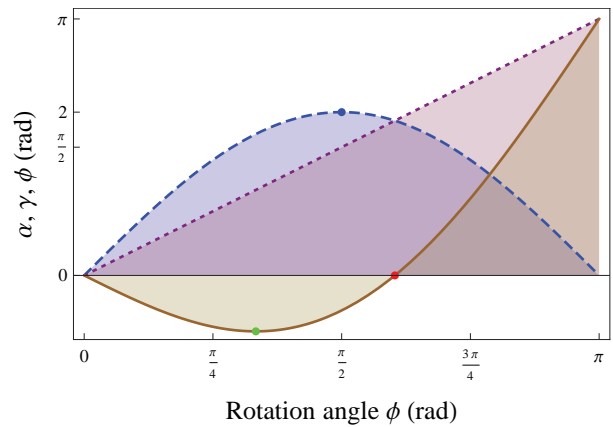}
\caption{\label{fig:2}(Color online) The dependence of dynamic and geometric phase components on the rotation angle (also the overall phase) $\phi$. The dashed (blue), dotted (purple) and solid (brown) curves correspond to the dynamic component $\alpha$, overall phase $\phi$ and the geometric component $\gamma$, respectively. The blue, red and green dots mark the maximum of dynamic phase, zero point and minimum of geometric phase respectively.}
\end{figure}

First we analyze the dependence of geometric and dynamic phase components on the spin rotation angle (also the overall phase) $\phi$. By the method utilizing the analytic solution introduced in Eq.~(\ref{psi}), numerical results are obtained and showed in Fig.~\ref{fig:2}. The dynamic phase $\alpha$ is a concave function of $\phi$, which is symmetrical to the axis of $\phi=\pi/2$. $\alpha$ is nonzero and positive except when $\phi=0,~\pi$. The maximum of $\alpha$ is 2 (blue dot in Fig.~\ref{fig:2}), which is in reasonable agreement with the theoretical expression of Eq.~(\ref{dynamic1}) equal to $2$ when $\phi=\pi/2$, $\Omega=\Delta$. The physical origin of the maximum dynamic phase is the instantaneous expectation value of the Hamiltonian $\bra{\psi(t)}H\ket{\psi(t)}$ attains maximum when system is in the eigenstate $\ket{\psi}=1/\sqrt{2}(\ket{z}+\ket{\tau})$, which can obtained when $\Omega=\Delta$. Removing the dynamic component from the overall phase leads to the geometric component $\gamma$, which is a convex function of $\phi$. On some interval $\alpha>\phi$, resulting in negative values of $\gamma$. Fig.~\ref{fig:2} shows the geometric component has a minimum $\sim-0.68$ when $\phi\approx1.05$ (green dot). When $\phi\approx1.90$ the overall phase is equal to the dynamic phase, leading to the zero value of geometric component (red dot). The concavity of geometric component indicate that only when the spin rotated by $\phi=\pi$, the overall phase is purely the geometric phase. For other rotation angles, $\phi$ consists of a nonzero dynamic component.

\begin{figure}[htbp]
\includegraphics[width=8cm]{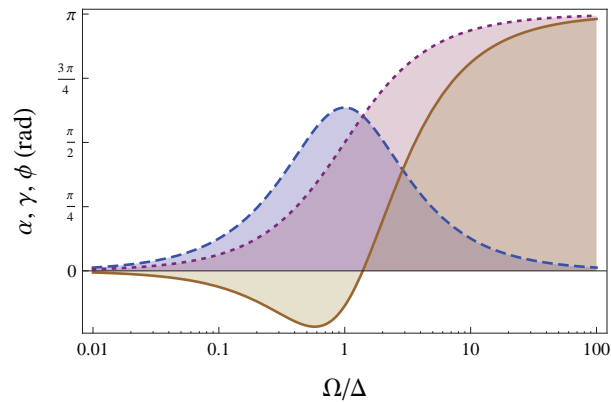}
\caption{\label{fig:3}(Color online) Log-linear plot of dynamic and geometric phase components as a function of the ratio of the Rabi frequency to the detuning, $\Omega/\Delta$. The dashed (blue), dotted (purple) and solid (brown) curves correspond to the dynamic component $\alpha$, overall phase $\phi$ and the geometric component $\gamma$, respectively.}
\end{figure}

As the review in Sec. II, the rotation angle is determined by $\Omega/\Delta$, therefore the phase components are essentially dependent on $\Omega/\Delta$. Fig.~\ref{fig:3} shows the log-linear plot of overall phase and two phase components as a function of $\Omega/\Delta$. The dynamic phase component still has a symmetry which is characterized more clearly. An interesting feature is that $\Omega/\Delta$ and its reciprocal leads to the same value of dynamic component, e.g. $\alpha(10)=\alpha(0.1)$, resulting from the feature of integrand in Eq.~(\ref{dynamic1}). The minimum and zero value of geometric component are obtained when $\Omega/\Delta\approx0.58$ and $\Omega/\Delta\approx1.39$, respectively.

\begin{figure}[htbp]
\includegraphics[width=8cm]{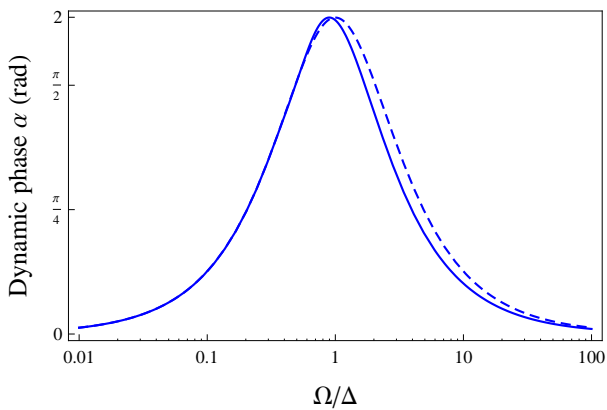}
\caption{\label{fig:4}(Color online) Log-linear plot of dynamic phase component as a function of $\Omega/\Delta$ by two methods. The dashed and solid curves correspond to the result by method I and method II, respectively.}
\end{figure}

Fig.~\ref{fig:4} shows the discrepancy of dynamic phase component by two methods. The solid curve corresponds to the result obtained by method II and the dashed curve by method I which is as the same as shown in Fig.~\ref{fig:3}. Both results are similar and in agreement with each other on distribution and variation trend. Especially when $\Omega/\Delta\ll1$ two curves basically coincide. Around $\Omega/\Delta=1$ the discrepancy appears. The dynamic phase by method II reaches its maximum earlier than the result by method I. With the increase of $\Omega/\Delta$, the dynamic phase by method II drops faster than the result by method I. But when $\Omega/\Delta\gg1$ two curves approach to each other again. The physical origin of the discrepancy is obviously the spin precession, which is neglected by method I but considered by method II. Due to spin precession about $x$ axis during the pulse action, a little population is transferred from $\ket{z}$ to $\ket{\bar{z}}$. This leads to not only degrading the unitarity of the spin rotation operation \cite{Economou07}, but also the change of system energy distribution. The trion state $\ket{\tau}$ has a higher energy than two lower states $\ket{z}$ and $\ket{\bar{z}}$. The population transferred to $\ket{\bar{z}}$ will not be transferred back to $\ket{z}$ during the pulse action because of $\omega_{B}\ll\Omega$, and $\ket{\bar{z}}$ will not be excitated to the trion state due to selection rules. The due population of the trion state during the pulse action is thus decreased. Therefore the contribution of the trion state to the expectation value of the Hamiltonian is lower than that for the no spin precession case. This leads to the discrepancy of dynamic phase. The reason for the discrepancy becoming more evident when $\Omega>\Delta$ is that more real population is transferred to the trion state during the pulse action. Thus the population transferred out of the two-level system results in more loss of contribution of the trion state to the dynamic phase.

\section{Elimination of dynamic phases by multipulse control}
\begin{figure}[htbp]
\includegraphics[width=8cm]{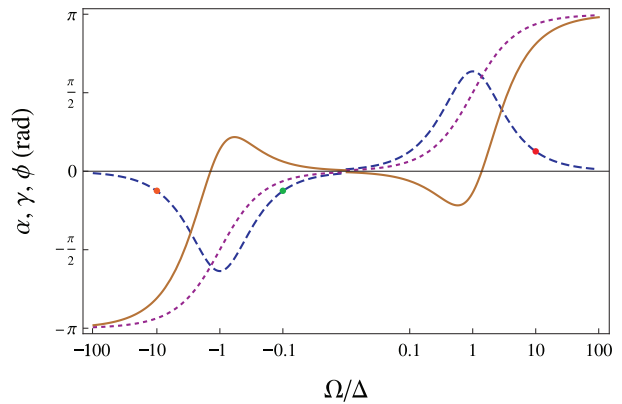}
\caption{\label{fig:5}(Color online) Log-linear plot of dynamic and geometric phase components as a function of $\Omega/\Delta$. The dashed (blue), dotted (purple) and solid (brown) curves correspond to the dynamic component $\alpha$, overall phase $\phi$ and the geometric component $\gamma$, respectively. Negative values of $\Omega/\Delta$ lead to negative overall and dynamic phases. The red, orange and green dots mark the values of dynamic phase when $\Omega/\Delta=10$, $-10$ and $-0.1$, respectively. Phases around $\Omega/\Delta=0$ approach zero and are thus not sketched.}
\end{figure}

Now we demonstrate the scheme to eliminate the dynamic phase by several $2\pi$ sec pulses control under slow precession approximation. Besides the symmetry of the dynamic phase to $\Omega/\Delta=\pm1$, the overall phase and phase components are all odd functions of $\Omega/\Delta$, as shown in Fig.~\ref{fig:5}. Utilizing these symmetry relations, we can choose particular parameters for each sec pulse, thus the dynamic phases induces by different pulses may cancel out. For the case of two pulses, we choose $r_{1}=\Omega/\Delta$ for pulse 1, e.g. the red dot in Fig.~\ref{fig:5}. For pulse 2, if $r_{2}=-r_{1}$, the dynamic phase is the opposite number of that induced by pulse 1, $\alpha(-r_{1})=-\alpha(r_{1})$ (e.g. the orange dot in Fig.~\ref{fig:5}). However, due to the odevity, the overall phase induced by pulse 2 is also the opposite number of that by pulse 1, i.e. $\phi(-r_{1})=-\phi(r_{1})$, $\gamma(-r_{1})=-\gamma(r_{1})$. Therefore after these two pulses all the phases cancel out. For the purpose of only eliminating the dynamic phases, we choose $r_{2}=-1/r_{1}=-\Delta/\Omega$ for pulse 2. For this case the dynamic phase is still the opposite number of that induced by pulse 1 due to the symmetry (e.g. the green dot in Fig.~\ref{fig:5}), but the overall phases do not cancel out, leading to the geometric phase being added. Therefore the total geometric phase after two pulses reads
\begin{eqnarray}
\gamma_{tot}=\phi_{1}+\phi_{2}=2\arctan(\frac{\Omega}{\Delta})+2\arctan(-\frac{\Delta}{\Omega})\label{total}.
\end{eqnarray}

\begin{figure}[htbp]
\includegraphics[width=8cm]{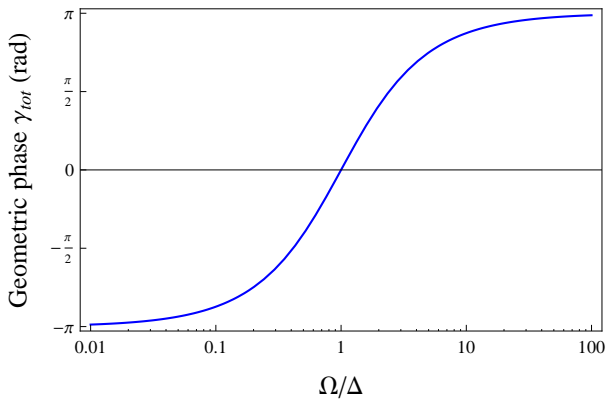}
\caption{\label{fig:6}(Color online) The total geometric phase $\gamma_{tot}$ obtained after two pulses, as a function of $\Omega/\Delta$. $r=\Omega/\Delta$, $-\Delta/\Omega$ for pulse 1 and 2 respectively.}
\end{figure}

Fig.~\ref{fig:6} shows the total geometric phase $\gamma_{tot}$ obtained after two pulses. It indicates that by our scheme the dynamic phase can be eliminated and arbitrary pure geometric phase can be obtained. Thus the rotation angle around $z$ axis is a pure geometric phase. In experiments, one may fix the pulse bandwidth $\Omega$, and only adjust the detuning from the trion resonance $\Delta$ to satisfy the parameter relation $r_{2}=-1/r_{1}$ for the two pulses.

\section{Fidelity}
Fidelity is a direct measure to characterize how accurate a gate operation is implemented. The dominate mechanisms to deteriorate the fidelity of our nonadiabatic geometric rotation are the spin precession induced by the magnetic field and the population loss induced by the decay of the trion state. The former takes effect during the pulse action and between the two pulses while the latter functions only during the pulse action. The times of spin dephasing due to phonon and hyperfine coupling are on the order of msec \cite{Khaetskii01} and $\mu$sec \cite{Khaetskii03} respectively and thus could be neglected. Here we consider the effects of spin precession and population loss, and calculate the average fidelity by numerical solution of Schr\"{o}dinger equation. The fidelity is defined as $\mathcal{F}(U)=\overline{|\bra{\Psi}U^{\dag}U_{id}\ket{\Psi}|^{2}}$, where $\ket{\Psi}$ is the initial state, $U_{id}$ and $U$ are the ideal and actual operations respectively, and the average is taken over all input states \cite{Bowdrey02}. We define $I=U^{\dag}U_{id}$, same as Ref.~\cite{Economou06,Piermarocchi02}, then the fidelity is derived as
\begin{eqnarray}
\mathcal{F}(U)=\frac{1}{3}\sum_{i}|I_{ii}|^{2}+\frac{1}{6}\sum_{i\neq{j}}(|I_{ij}|^{2}+I_{ii}I_{jj}^{*})\quad(i,j=1,2),
\end{eqnarray}
where $I_{ii},~I_{ij}$, and $I_{jj}$ are matrix elements of the operator $I$.

In calculation, the actual operation $U$ is firstly taken over the 3-dimensional Hilbert space then truncated in the $\{\ket{\bar{z}},~\ket{z}\}$ subspace. Due to the effect of population loss, $U$ is nonunitary. We take the parameters realistic in experiments as follows: the magnetic field $B=0.29$T \cite{Greilich09} and the electron $g$ factor $|g_{e}|=0.57$ \cite{Greilich06}, which lead to the Larmar period $T\thickapprox0.43$ns, the pulse duration $\tau_{d}=1.5$ps \cite{Greilich09}, and the trion lifetime $\tau_{t}=900$ps \cite{Ware05}. The two pulses are successively applied and the distance between pulse centers is taken as $14\tau_{d}$. We notice in Ref.~\cite{Schirmer05} the square-wave pulse is adopted to realize geometric phase gates for QD charge qubits and the imperfection of the pulse shape is taken into account because it results in inaccurate pulse area. Here we consider the two pulses with a good hyperbolic secant shape and the pulse area is independent of the detuning \cite{Economou07}. In experiments, for the geometric rotation of $\pi$, only one pulse is required because it leads to a pure geometric phase, as demonstrated in Sec. III. The fidelities of other selected rotations of spin by multipulse control for studied QD parameters when $B=0.29$T are listed in Table I.

\begin{table}
\caption{Fidelity of selected nonadiabatic geometric rotations about $z$ for
considered QD parameters when $B=0.29$T.}
\begin{ruledtabular}
\begin{tabular}{cc}
\centering
$\gamma_{tot}$ (rad)&Fidelity\\
\hline
$\pm\pi/4$ & $99.38\%$ \\
$\pm\pi/2$ & $99.43\%$ \\
$\pm3\pi/4$ & $99.48\%$ \\
\end{tabular}
\end{ruledtabular}
\end{table}

\begin{figure}[htbp]
\includegraphics[width=8cm]{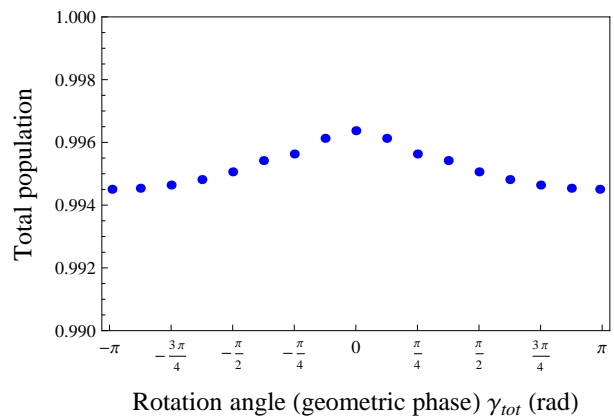}
\caption{\label{fig:7}(Color online) Total population after two pulses, as a function of the rotation angle for studied QDs. The transverse magnetic field is taken as $B=8$T.}
\end{figure}

\begin{figure}[htbp]
\includegraphics[width=8cm]{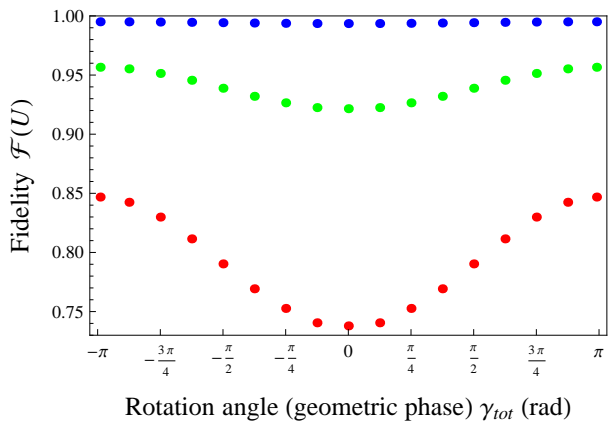}
\caption{\label{fig:8}(Color online) Fidelity of the operation
as a function of the rotation angle for studied QDs. The upper (blue), medium (green) and lower (red) dots correspond to the fidelities when the transverse magnetic field is taken as $B=0.27$T, $1.35$T and $2.7$T, respectively.}
\end{figure}

The numerical results show the fidelities are equal for the positive and negative rotation angles which have the same absolute value. This is in full agreement with the analysis result in Eq.~(\ref{total}): for $\pm\gamma_{tot}$ only the sign of the detuning $\Delta$ is changed, the total stimulating effect of the two pulses are thus the same. The loss of fidelity is due to the combined action of the spin precession and population loss. Small angle corresponds to the detuning $\Delta$ close to $\Omega$ for both pulses. For this case the effect of spin precession is predominante due to the majority of virtual trion excitation. Large angle corresponds to one pulse with small $\Delta$ while the other pulse with large $\Delta$. For this case the effect of population loss is enhanced (see Fig.~\ref{fig:7}) due to the increased real trion excitation. Lower fidelity for smaller angle indicates that the effect of spin precession is stronger than population loss, which is more clearly shown in Fig.~\ref{fig:8}. When larger magnetic field is applied, the fidelity of small angle decreases more significantly than that of large angle. This indicates that compared with decay of trion state, the effect of spin precession is more detrimental to the fidelity, and weak magnetic field is a crucial condition for high fidelity of geometric rotation.

\section{Summary and prospects}
In summary, we briefly review the fast spin rotation proposal proposed by  Economou \textit{et al}. \cite{Economou06} and analyze the geometric and dynamic phase components of the overall phase induced by $2\pi$ sec pulses. The dependence of two phase components on the ratio of the Rabi frequency to the detuning from the trion resonance is investigated. Numerical results indicate that if one pulse is applied, the overall phase is purely the geometric phase only when the spin is rotated by $\pi$. With other rotation angles the dynamic component is nonzero. An interesting effect is shown that the dynamic phase has a symmetry as a function of rotation angle and $\Omega/\Delta$. Results also show the discrepancy of dynamic phase for the cases with and without slow precession approximation. We discuss its physical origin: the effect of spin precession. The spin precession leads to population loss out of the two-level system during the pulse action, resulting in the lower contribution of the trion state to the system energy. Utilizing the symmetry relations of the phases, we propose a scheme to eliminate the dynamic phase by multipulse control. By choosing the ratio $\Omega/\Delta$ as the negative reciprocal to each other for each pulse, the dynamic phases induced by two pulse cancel out. Considering spin precession and population loss, high fidelity of the geometric rotation is obtained when the weak magnetic field is applied. The detrimental effects of magnetic field and decay of trion state are compared and discussed. Our scheme realizes high fidelity of geometric rotation of electron spin about an axis. Because of the evident advantage on short manipulation time and considerable feasibility of the proposal $\mathcal{T}$ \cite{Greilich09}, our scheme may generalize the proposal to universal nonadiabatic GQC. Further research may be focused on nonadiabatic two-qubit geometric gates in the quantum dot system.

\begin{acknowledgements}
We thank Hao-Di Liu for valuable discussions. This work is supported by National Natural Science Foundation of China (NSFC) under grant no 10875020 and 60703100.\\
\end{acknowledgements}

\end{document}